\documentclass[a4paper,10pt,twoside]{cpc-hepnp}

\usepackage{multicol}
\usepackage{graphicx}
\usepackage{booktabs}
\usepackage{amssymb,bm,mathrsfs,bbm,amscd}
\usepackage[tbtags]{amsmath}
\usepackage{lastpage}

\begin{document}

\fancyhead[co]{\footnotesize NA Xue-Sen~et al: A new parametric equation of state and quark stars}

\footnotetext[0]{Received 22 September 2010}

\title{A new parametric equation of state and quark stars \thanks{Supported by China Scholarship Council, NSFC(10973002, 10935001) and the National Basic Research Program of China (grant 2009CB824800). }}

\author{%
      NA Xue-Sen $^{1)}$\email{naxuesen@pku.edu.cn}%
 and XU Ren-Xin $^{2)}$\email{r.x.xu@pku.edu.cn}%
}
\maketitle

\address{%
School of Physics and State Key Laboratory of Nuclear Physics and Technology, Peking University,
Beijing 100871, China\\
}

\begin{abstract}
It is still a matter of debate to understand the equation of state of cold supra-nuclear matter in compact stars because of unknown non-perturbative strong interaction between quarks. Nevertheless, it is speculated from an astrophysical view point that quark clusters could form in cold quark matter due to strong coupling at realistic baryon densities. Although it is hard to calculate this conjectured matter from first principles, one can
expect the inter-cluster interaction to share some general features to nucleon-nucleon interaction
successfully depicted by various models. We adopt a two-Gaussian component soft-core potential with
these general features and show that quark clusters can form stable simple cubic crystal structure
if we assume the wave function of quark clusters have a Gaussian form. With this parameterizing, Tolman-Oppenheimer-Volkoff equation is then solved
with reasonable constrained parameter space to give mass-radius relation of crystalline solid quark
star. With baryon densities truncated at $2n_0$ at surface and
range of interaction fixed at $2$fm we can reproduce similar mass-radius relation to that obtained with bag model equations of state. The maximum mass ranges from $\sim 0.5M_\odot$ to $\gtrsim 3M_\odot$.
Observed maximum pulsar mass ($\gtrsim 2M_\odot$) is then used to constrain
parameters of this simple interaction potential.
\end{abstract}



\begin{multicols}{2}

\section{Introduction}
An understanding of cold quark matter is both one of the most challenging problem in
particle physics and a prerequisite to understand the true nature
of pulsars and pulsar-like objects. However, due to both non-perturbative
nature of strong interaction at low energy and complexity presented
by the quantum many-body problem, it is almost impossible to understand
such state theoretically from first principles.

Over the decades various approaches to bypass these difficulties has been developed,
both perturbatively such as that of color super-conductivity~\cite{alford08} and non-perturbatively such as
lattice QCD and QCD-based effective models. On the other hand, it has been conjectured~\cite{xu03}
that quark matter could be in a solid state at extreme low temperature present
in pulsar interior. This possibility could combine naturally with several previous works suggesting
the possibility that deconfined quark matter might contain quark clusters of $3N$ valence
quarks~\cite{clark86,schulz87,bpzszp88} into a reasonable conjecture that quark clusters
could form crystal lattices.

Because of the difficulty to obtain detail of the interaction between quark clusters and therefore equation of state
of cold quark matter at a few times nuclear densities,
it is interesting to apply simple phenomenological models. If we can use astronomical
observations to constrain parameters in such models, we will be able to gain some insight on
properties of low-energy QCD or rule out such form of cold quark matter within pulsars.

In several previous works (\cite{xylai09Poly},\cite{xylai09LJ}), different models have been tried to investigate the possible equation of state of solid quark matter and have provided possibility to explain stiffness in equation of state required by observed massive pulsars~\cite{massiveNS08}.
In \cite{xylai09LJ} the Lennard-Jones potential which was introduced to model interaction between inert gas
molecules~\cite{LJ} is used as potential between two quark clusters. The Lennard-Jones potential shares some basic properties with nuclear forces such as short-range repulsion and longer range attraction (c.f.~\citep{kranebook88}).
In this article we adopt a more realistic parametrization that shares great similarity with various models
depicting hyperon-hyperon potential. It has been shown that
the interaction between two H-dibaryons -- cluster of 6 valence quarks may also share this general feature~\citep{sakai97}.

This article is arranged as follows. The model of inter-cluster potential is presented in Section 2.
Parameter space used for calculation is discussed in Section 3. Section 4 shows result of calculation.
Conclusion and some discussions are presented in section 5.

\section{Inter-cluster potential}
As in \cite{xylai09LJ} we consider quark clusters
with $3N$ valence quarks with $N=1,6$ and a simple cubic lattice structure is adopted for simplicity.
In the context of strange quark matter, these are particles
with the same valence quark composition as hyperons and the hypothetical `quark-alpha'~\citep{michel88}.
By extension of the Lennard-Jones potential, we adopt the following simple parametrization for effective interaction between
two quark clusters localized on crystal lattice sites,
\begin{equation}
v(r)=V_1e^{-\left(\frac{r}{r_1}\right)^2}-V_2e^{-\left(\frac{r}{r_2}\right)^2},
\end{equation}
where possible spin-dependent interactions are omitted for simplicity.
As is mentioned above, with the condition $V_1>V_2$, $r_1<r_2$, this potential qualitatively
reproduces the general feature of various successful phenomenological potentials
of nuclear interactions: soft-core repulsion at short range and attraction at longer range.

It turned out that maximum mass and mass-radius curve is
not sensitively depend on the value of $r_2$. Therefore
it is reasonable to fix $r_2$ at 2fm which is a typical range of nuclear force.

For simplicity we assume that cluster center-of-mass has a Gaussian wave packet with width $w$
as wave function
\begin{equation}
\psi_{\boldsymbol{r}_0,w}(\boldsymbol{r})=\frac{1}{\pi^{3/4}w^{3/2}}e^{-\frac{\vert \boldsymbol{r}-\boldsymbol{r}_0 \vert^2}{2w^2}},
\end{equation}
In~\cite{xylai09LJ} it is assumed that potential well created by surrounding clusters
with Lennard-Jones interaction is deep enough to trap quark clusters in the potential well.
To ensure that soft-core potential with given parameter can also achieve this we
adopt variational method, i.e. to determine the value of $w$ by minimizing total energy of a single cluster which is a sum of kinetic energy of the wave packet and potential energy
contributed by surrounding cluster lattices.
The result shows that with the range of
of parameters considered in this work, the width of wave packet is rather small
compared to inter-cluster distance and hence it makes sense to
speak of this system of clusters as quark clusters trapped in periodic lattice.
With this small width, the overlap between adjacent wave packets is negligible. Thus
it is reasonable to omit the difference
between fermionic and bosonic quark clusters.

To calculate total contribution to single cluster potential energy, a sum is taken
over a cube of $21^3$ lattices centered around the quark cluster under consideration.
The size of the cube is enough since cluster number density in this work will
not exceed $\sim 10 n_0$. Hence, twice the total contribution
to potential energy for a single cluster is
\end{multicols}
\ruleup
\begin{equation}
\label{one}
V(n)\equiv \left(\sum_{k_1=-10}^{10}\sum_{k_2=-10}^{10}\sum_{k_3=-10}^{10}\right)' \tilde{v}\left(w;\frac{\sqrt{k_1^2+k_2^2+k_3^2}}{n^{1/3}}\right),
\end{equation}
\ruledown \vspace{0.5cm}
\begin{multicols}{2}
\noindent where the prime means that the sum omits $k_1=k_2=k_3=0$, and
\begin{equation}
\tilde{v}(w;r)=\int d^3\boldsymbol{r}'\psi^*_{\textbf{0},w}(\boldsymbol{r}')v(r')\psi_{\boldsymbol{r},w}(\boldsymbol{r}'),
\end{equation}
is the expectation value of potential energy between two clusters.

The total energy density is then
\begin{equation}
\epsilon=\frac{n}{2}V(n,w)+nm+\frac{3}{4mw^2}+\frac{9}{8}(6\pi^2)^{1/3}\hbar vn^{4/3},
\end{equation}
where the third term comes from contribution of kinetic energy
and $w$ is treated as a function of number density $n$. Same as in~\cite{xylai09LJ} the fourth term
comes from zero-point energy of phonon in Debye's approximation, with
\begin{equation}
\frac{1}{v^3}=\frac{1}{3}\left(\frac{1}{v_\parallel^3}+\frac{2}{v_\perp^3}\right),
\end{equation}and $v_\parallel,v_\perp$ stands for sound velocity of longitude
and transverse modes respectively. Following the argument of~\cite{xylai09LJ}
we take $v=c/3$ because its value have only very small influence on the final result.

Ignoring effect of temperature the pressure can be derived as
\begin{align}
P=n^2\frac{d}{dn}\left(\frac{\epsilon}{n}\right)=\frac{n^2}{2}\frac{\partial V}{\partial n}+\frac{n^2}{2}\frac{\partial V}{\partial w}\frac{dw}{dn}-\frac{3}{2mw^3}\frac{dw}{dn}.
\end{align}
With energy density and pressure one can establish equation of state
and solve Tolman-Oppenheimer-Volkoff (TOV) equation with varying central density
to get the mass-radius relation.
In practice, it is more convenient to skip (numerical) determination of equation of state
and write the TOV equation in terms of $n(r)$ and $M(r)$,
\end{multicols}
\ruleup
\begin{align}
\frac{dn}{dr}&=-\frac{G}{r^2-2GMr}(P(n)+\epsilon(n))(M+4\pi r^3P(n))\left(\frac{dP}{dn}\right)^{-1}\\
\frac{dM}{dr}&=4\pi r^2\epsilon(n)\\
M(0)&=0, \quad n(0)=n_c
\end{align}
\ruledown \vspace{0.5cm}
\begin{multicols}{2}
\noindent where $n_c$ is central number density of quark cluster.
Because quark matter is usually expected to be self-bound at zero pressure (for instance
under bag model equation of state),
it is reasonable to simply adopt a truncation baryon number density $n_{\text{surf}}$ at the surface. In this
work we adopt $n_{\text{surf}}=2n_0$ where $n_0$ is baryon number density of normal nuclear matter.

\section{Parameter space}
As is stated above, $r_2$ is fixed to $2$fm. Therefore, we have 4 free parameters: quark cluster mass $m$; height of the two Gaussians in the potential $V_1,V_2$; range of repulsive core $r_1$.
It is appropriate to expect that the depth of attractive part of the potential might be of the same order of magnitude
as typical potential between two nucleons in nuclear matter.
Therefore we fix $V_2$ at $50$MeV and $100$MeV respectively. On the other hand, we fix the value of $m$ to $1$GeV and $6$GeV for 3-quark clusters and the aforementioned hypothetical `quark-alpha'. Here mass of 3-quark cluster is take from mass of $\Lambda$ hyperon ($1115$MeV). The 18-quark cluster -- the `quark-alpha' is assumed to contain 6 quarks of each flavor which may have a mass less than $6m_\Lambda$ but we omit this possibly small difference for simplicity. Thus all that are left is $V_1$ and $r_1$. We adopt a condition,
\begin{equation}
V_1r_1^3>V_2r_2^3,
\end{equation}which ensures that potential energy is always positive (i.e., repulsive) when density is very high.

\section{Results and conclusions}
With the above settings of parameter space, we drew 4 contour plots of maximum mass calculated for 4 different sets of $(m,V_2)$ which are shown in Fig.\ref{fig1}.

\begin{center}
\includegraphics[width=8cm]{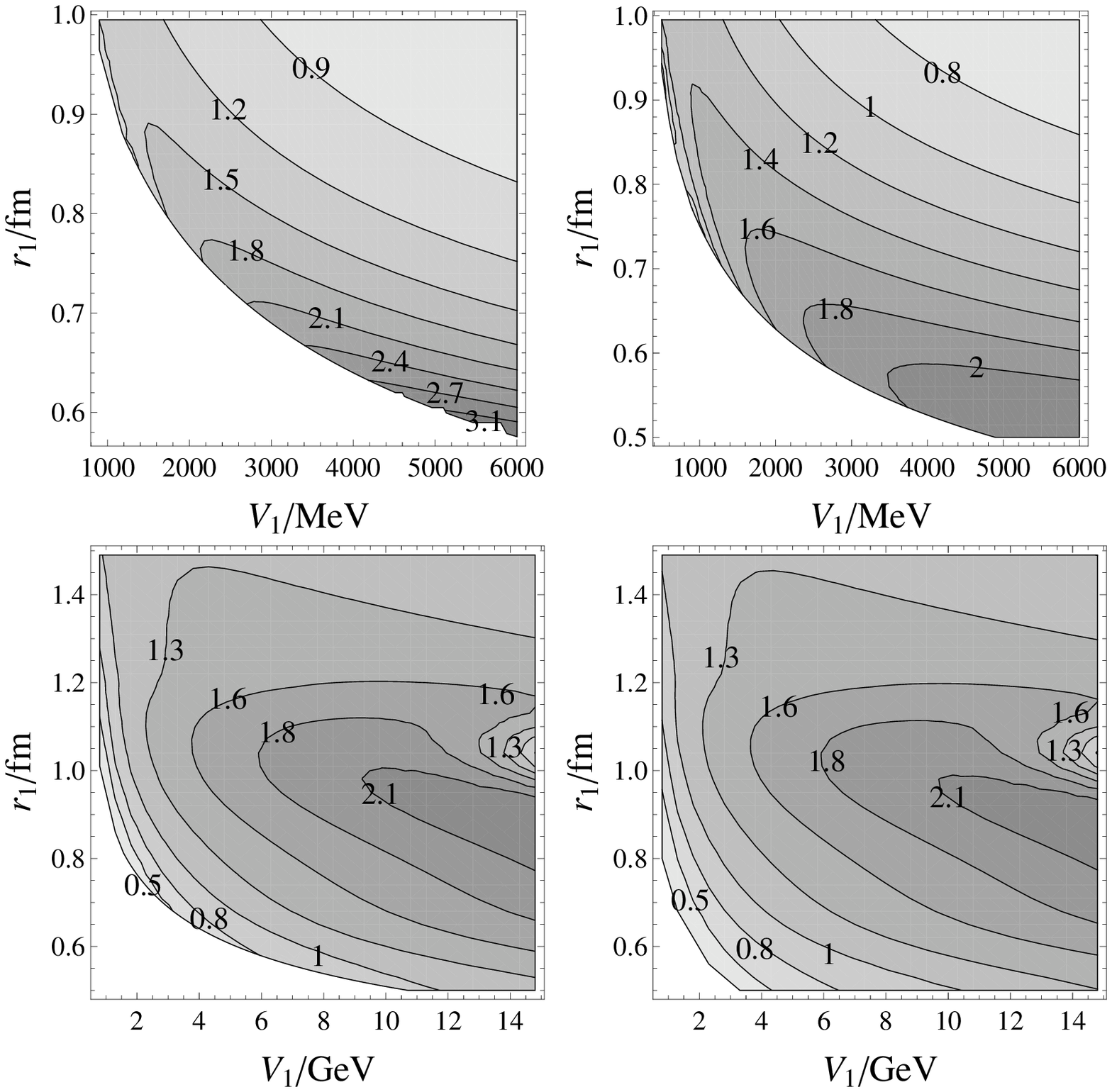}
\figcaption{\label{fig1} Maximum mass in unit of solar mass of solid quark star. \emph{Upper left}: ~$m=1$GeV, $V_2=100$MeV; \emph{Upper right}: $m=1$GeV, $V_2=50$MeV; \emph{Lower left}: $m=6$GeV, $V_2=100$MeV; \emph{Lower right}: $m=6$GeV, $V_2=50$MeV. Boundary of contour plots are moved a little upper right than $V_1r_1^3=V_2r_2^3$ to avoid parameters that lead to large error in numerical calculations}
\end{center}

Typical mass-radius relation curves of these settings with maximum mass exceeding $1.9 M_\odot$ are also shown (Fig.\ref{fig2}). From Fig.\ref{fig1} we can see that for our simple soft-core parametrization, maximum mass can range roughly from below $1M_\odot$ to about $\sim 3M_\odot$ for solid quark stars with 3-quark cluster forming crystal lattice and from below $0.5M_\odot$ to about $2.1M_\odot$ for solid quark stars made up of `quark-alpha' particles. On the other hand typical mass-radius relations shown in Fig.\ref{fig2} are very similar to those calculated within bag model EoS (shown as gray curves in Fig.\ref{fig2}, adopted from model SS1, SS2 in~\cite{SS1SS2}) and M-R curves calculated in \cite{xylai09LJ}. This shows that at least for some region of parameter space our simple parametrization can also produce heavy maximum mass supported by observed value of $2M_\odot$. Inversely, with current observation we can already restrict parameters in this very simple model with only 4 parameters. For instance, to get maximum mass larger than $1.9M_\odot$ for $m=1$GeV and $V_1<6$GeV, we have to restrict $r_1$ to below $0.75$fm when $V_2=100$MeV and restrict $r_1\lesssim 0.6$fm and for $m=6$GeV it requires $r_1\lesssim 1$fm and $V_1 \gtrsim 8$GeV.

\begin{center}
\includegraphics[width=7.5cm]{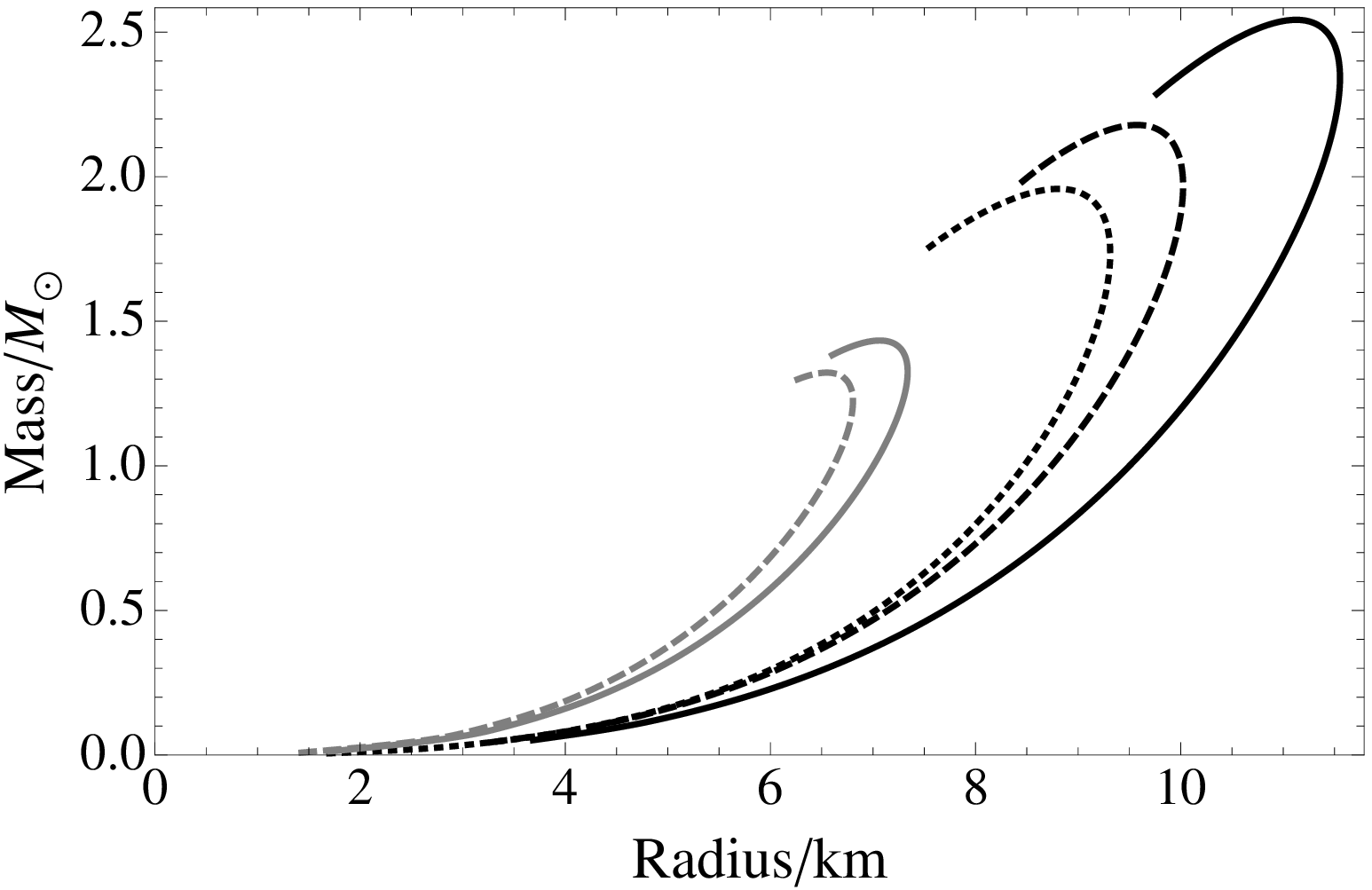}
\figcaption{\label{fig2} Mass-radius relation curves for typical parameters giving large maximum masses.  \emph{Solid}: $m=1$GeV, $V_1=3.2$GeV, $V_2=100$MeV, $r_1=0.68$fm; \emph{Dashed}: $m=1$GeV, $V_1=5$GeV, $V_2=50$MeV, $r_1=0.54$fm, \emph{Dotted}: $m=6$GeV, $V_1=9$GeV, $V_2=100$MeV, $r_1=0.95$fm; \emph{Gray solid and dashed}: Bag model EoS SS1 and SS2 adopted from~\citep{SS1SS2} }
\end{center}

\section{Discussion}
In cold quark matter at baryon number densities realistic for compact stars, the interaction between quarks could be strong enough that instead of condensating in momentum space to form color-superconductive phase it is possible that dressed quarks undergo condensation in position space to form quark clusters. As is stated in \cite{xylai09LJ} if the potential well formed by neighboring clusters are deep enough to trap quark cluster, cold quark matter could form crystal solid in low temperature.

In this work we discussed simple-cubic lattice structure formed by 3 and 18-quark clusters using a simple two Gaussian component parameterization of soft-core potential to simulate the interaction between quark clusters. This parametrization shares the basic properties of nucleon-nucleon interaction mediated by meson exchange -- short range repulsion, medium and long range attraction and a finite range. These properties are also shared by Lennard-Jones potential adopted in \cite{xylai09LJ}. However unlike Lennard-Jones potential with $r^{-12}$ pole at origin, soft-core potential adopted in this work can be treated by non-relativistic quantum mechanics. By minimization total energy we found that at realistic densities this soft-core potential can lead to a stable lattice structure. It is entirely possible that other unit cell structure (e.g. body-centered cubic) is more stable, but we expect the difference to be quantitative instead of qualitative.

It is also worth mentioning that despite great similarity between mass-radius relation obtained with bag model equation of state and those calculated here, the underlying picture is drastically different. In bag model, quark star is degenerate Fermi gas of free quarks sustained by vacuum energy and associated negative pressure. In this work and \cite{xylai09LJ} pressure is mainly provided by repulsive core of inter-cluster potential instead of mere degenerate pressure.

Although maximum mass cannot be easily tuned to $\sim 6M_\odot$ as in \cite{xylai09LJ} due to soft-core nature of the interaction, our parametrization can still provide a stable lattice crystal structure with maximum mass exceeding $2M_\odot$ which is in accordance with observed maximum mass\cite{massiveNS08}. Inversely, the observed maximum mass can be used to put constraints on parameters of this simple model which will possibly give some insights into the form of interaction between quark clusters if such phase exists.

\acknowledgments{We thank pulsar group at Peking University for useful discussions.}

\end{multicols}

\vspace{105mm}

\vspace{-1mm}
\centerline{\rule{80mm}{0.1pt}}
\vspace{2mm}

\begin{multicols}{2}

\end{multicols}

\clearpage

\end{document}